\documentclass[11pt,twoside]{article}

\usepackage{asp2004}
\usepackage{epsf}
\usepackage{psfig}
\usepackage{graphicx}

\usepackage{lscape}

\begin{document}
\title{High spatial resolution study of the inner environment around two young planetary nebulae}   
\author{O.~Chesneau}   
\affil{Observatoire de la Côte d'Azur, Dpt. Gemini-CNRS-UMR 6203, Avenue Copernic, F-06130 Grasse}    

\begin{abstract} 
We present observations of the dusty emission from the young planetary
nebulae Hen\,2-113 and CPD-56 obtained with VLT/NACO, VLTI/MIDI. The central stars of these two objects are Wolf-Rayet stars of the same spectral type [WC10]. They share an impressive number of characteristics and are located at similar distance, making the detection of any differences in their close environment of great interest.
Hen\,2-113 exhibits a clear ring-like structure
of about 0.4 arcsec in diameter, superimposed to a more diffuse environment visible in L', M' and
8.7$\mu$m bands. No clear core could be detected for this object with MIDI through the N band.  
The dusty environment of CPD-56$^\circ$8032 is much more compact, dominated by
a bright, barely resolved, core whereas the visible nebula exhibits an amazing
complexity. From MIDI 8.7$\mu$m acquisition images (dominated by PAHs
emission), the extension and geometry of the core have been estimated and
compared to the STIS/HST observations (De Marco et al., 1997 and
2002). Moreover, high SNR fringes at low level have been detected with
projected baselines between 40 and 45 meters. This clear signal is interpreted
in terms of the bright inner rim of a dusty disk exposed to the flux from the Wolf-Rayet star. The geometrical parameters of the N band flux distribution are well constrained by means of simple geometrical models and a simple radiative transfer model has been developed to extract the physical parameters of the disk.

\end{abstract}


\section{Young PNs and B[e] stars}   
The optical spectrum of many
compact Planetary Nebulae (PNe) shows strong Balmer emission lines
and forbidden emission lines but we will concentrate our discussion on the strong IR
excess exhibited by these sources.
First of all, it is important to note that the targets discussed in this
paper, CPD-56$^\circ$8032 (He 3-1333, hereafter CPD) and Hen\,2-113
(He\,3-1044, hereafter HEN), are {\it not} PNe which could be confused with
B[e] star and classified as cPNB[e] stars or SymB[e] stars (Lamers et
al. 1998). Nevertheless, their study is highly relevant on the subject of this
workshop since the environment of these young PNe share many characteristics
encountered on B[e] stars ones. Among them, the most striking one is the
presence of an equatorial overdensity of dust material exhibited by a growing
number of PNs thanks to the improvement of high spatial resolution instruments
operating in the mid-IR domain. 

One of the most debated question regarding the post-AGB (asymptotic giant branch) evolution of low and
intermediate mass stars is the departure from spherical symmetry observed in
circumstellar envelopes of Pre-Planetary Nebulae (PPNe) and PNe. Indeed, whereas AGB stars have envelopes roughly spherical, many PNe
exhibit axisymmetric structures or even more complex morphologies.
Most of PNe shaping theories rely
on the velocity field of circumstellar material distributed either in an expanding toroidal 
structure, or in an equatorial disk. The formation of such structures requires the presence 
of a binary companion, rotation and/or magnetic fields.

We would like to mention the difficulty to classify  some B[e] sources in cPNB[e] stars or SymB[e] stars which reflects simply the difficulty to detect binary companion around post-AGBs and young PNe (see a recent example in S\'anchez Contreras et al. 2004). For instance, the PNe M\,2-9 and Mz\,3, classified as cPNB[e] in Lamers et al. (1998) show indirect evidence suggesting that they indeed harbor symbiotic systems (Smith et al. 2005, Smith \& Gehrz 2005).
This question of binarity is the core of the debate for the shaping of PNe since it is the most natural physical process able to explain the angular momentum stored in the nebulae (see Soker 1997, Soker \& Rappaport 2000, De Marco et al. 2004 and the recent provocative discussion in Soker 2005).

Wolf-Rayet  CSs of PNe , which represent about 10\% of all PNe,
are H-deficient stars (mostly [WC] type) that exhibit a dense stellar wind which shares many characteristics with is counterpart from higher mass population I WR. Among
the coolest CSs in this group are the [WC10] (Crowther, De Marco, \& Barlow 1998)
CPD and HEN.

The ISO spectrum of these objects is rich.
The Red Rectangle, CPD and HEN are
archetypal PAH-band sources  and have amongst the highest ratios of
PAH-band fluxes to total IR flux known. ISO spectroscopy also
found a chemical dichotomy with evidence of crystalline silicate at longer wavelengths (Waters
et al. 1998). The initial interpretation of the cold crystalline
silicates was that they probably were located at large distances
from the star, in order to be cool enough. However, the rate of occurrence 
of dual dust chemistry objects is now too high to
be consistent with objects that happened to be caught during
the brief transition from O-rich to C-rich chemistry. Binarity and
the presence of circumstellar disks or tori in which O-rich grains
from a previous evolutionary phase are increasingly
believed to be responsible for the dual dust chemistry (Cohen et al. 1999, 2002). 

De Marco \& Soker (2002) proposed that the correlation between Wolf-Rayet
characteristics and dual-dust chemistry points to a common mechanism within a
binary evolution framework. In their first scenario a low-mass main-sequence
star, brown dwarf, or planet spirals into the asymptotic giant branch star,
inducing extra mixing, hence a chemistry change, and terminating the
asymptotic giant branch evolution. In the second scenario a close binary
companion is responsible for the formation of a disk around either the binary
or the companion. This long-lived disk harbors the O-rich dust. These models
are mostly speculative owing to the rarity of spatial information on the dust
environment at the center of these systems and the scarcity of of binary
companion detection.

A careful study of the fundamental parameters of the CSs
and nebulae of HEN and CPD was conducted by De Marco et al. (1997, hereafter DBS97) and De Marco \& Crowther (1998).
HEN and CPD offer a unique opportunity to compare two different nebulae
ejected from central stars presenting striking similarities, especially in
terms of estimated distance (D=1.3-1.5\,kpc). This implies that the two stars
and their nebulae have been observed generally with the same instruments
involving similar spectral and spatial resolving power but the lack of spatial
resolution prevented De Marco and collaborator to perform an in-depth study of
the nebulae whereas their central stars  study has been exhaustive.

In the present work we report new observations of HEN and CPD performed with high spatial resolution techniques, namely adaptive optics with NACO/VLT and long baseline interferometry with MIDI/VLTI which allow us to observe with great detail
the bulk of the dust emission in the very central region of the object.

\begin{figure*}
  \begin{center}
\includegraphics[width=7.cm]{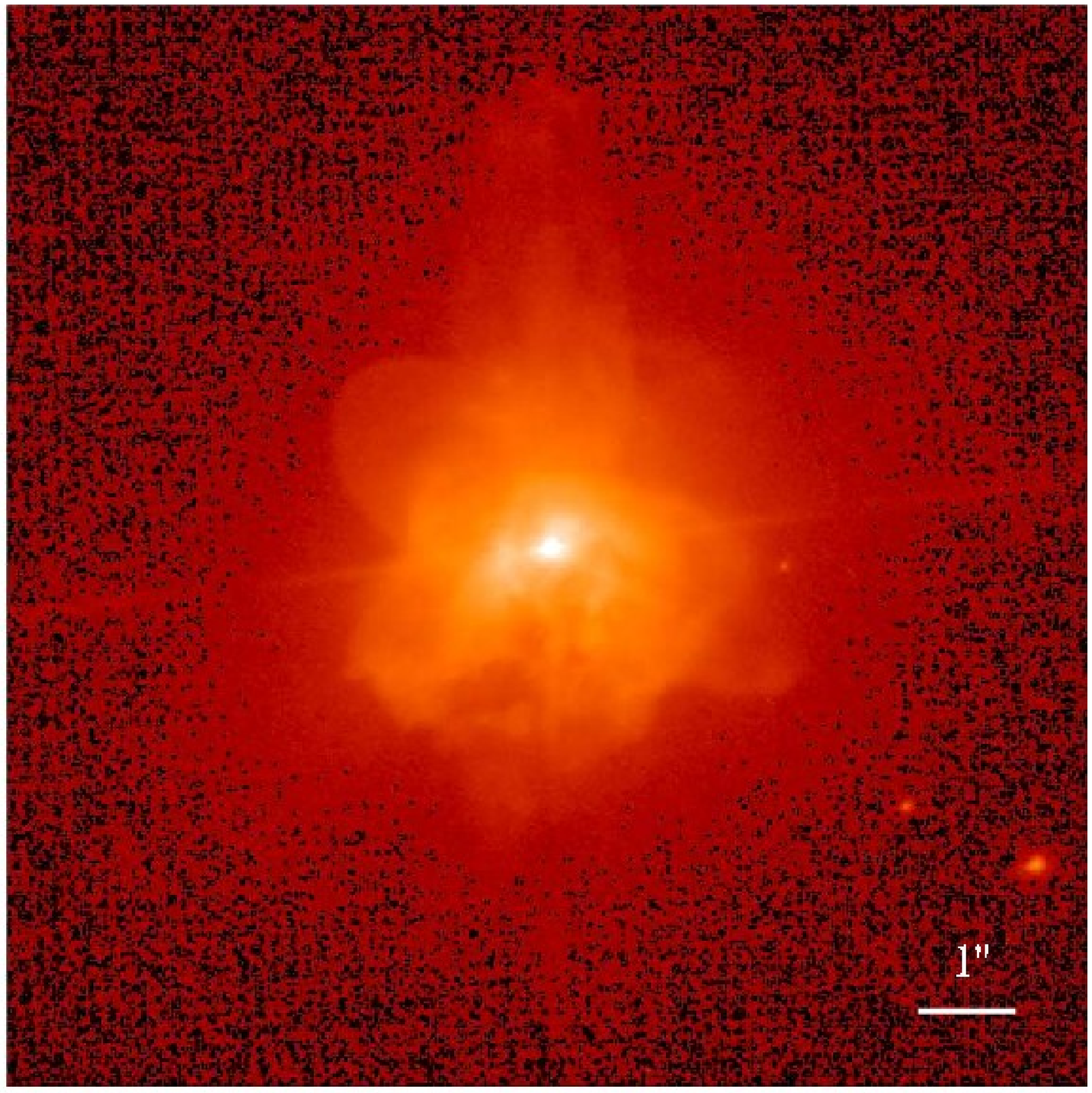}
\includegraphics[width=6.cm]{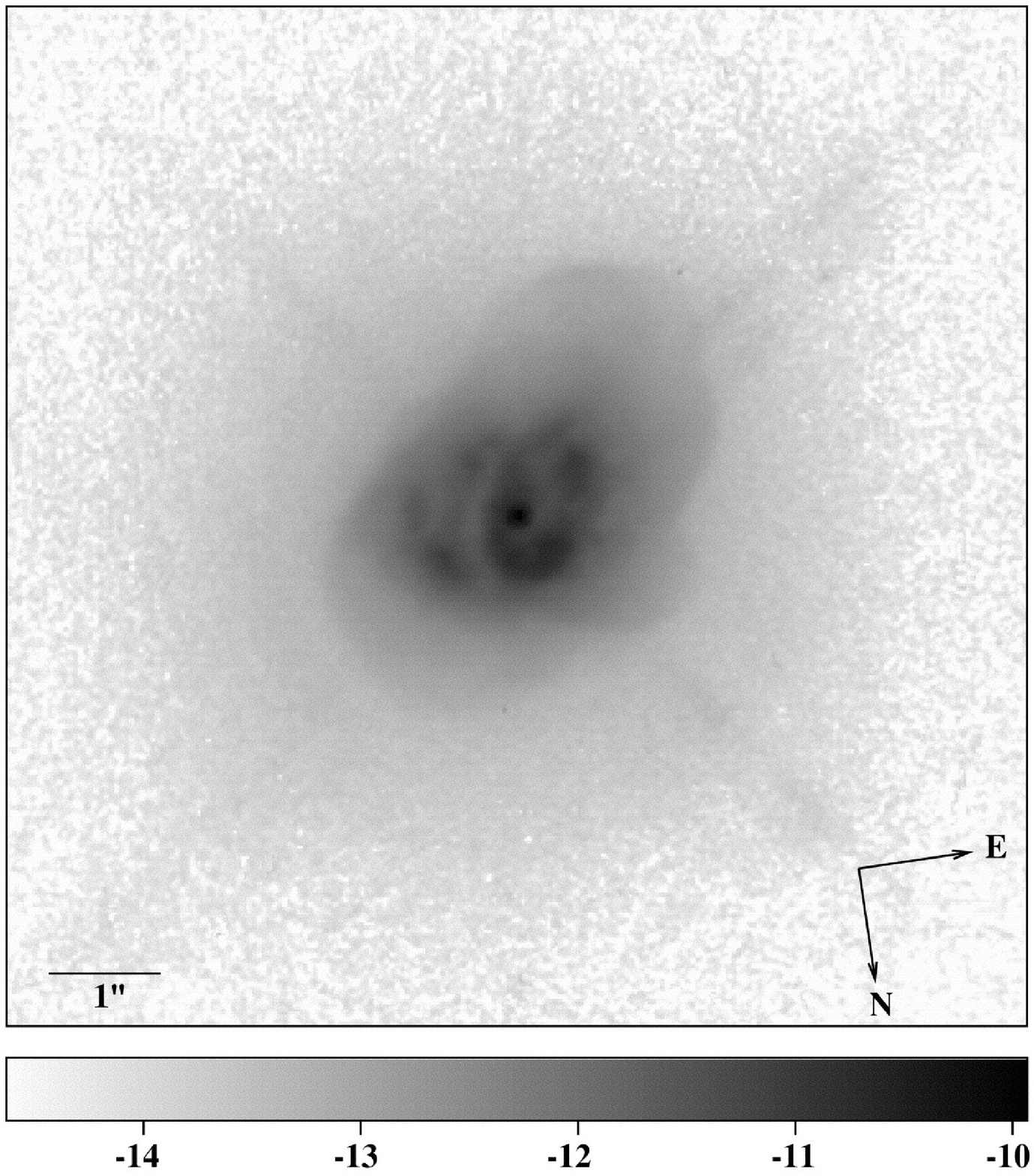}
  \end{center}
 \caption[]{F435 HST image of the young planetary nebula CPD-568032 with the Planetary Camera (resolution 0.0456$\arcsec$pixel$^1$) of WFPC2/HST . A logarithmic stretch has been used and North is up, East is left.
\label{fig:specfringes}}
\end{figure*}

\section{Hen\,2-113}
Hen 2-113 is a young planetary nebula with a [WC10] central star. Its ISO
spectra shows simultaneously the presence of C-rich (PAHs) and O-rich dust
grains. The distance to the star is in the range 1.5-3.1 kpc, Its central star has
$\rm T_{\rm eff}\sim 25 000$K and a mass-loss rate of $\sim~10^{-7}\rm
M_{\odot} \rm yr^{-1}$.

Despite the careful study of the fundamental parameters of the CSs
by De Marco and collaborators,
the morphology of the nebula of HEN was poorly known until
the HST observations by Sahai et al. (2000). These authors showed that
HEN exhibits a complex geometry, roughly bipolar with two bright,
knotty, compact ringlike structures around the central star. This compact structure is embedded in a larger and fainter spherically symmetric structure.
HEN was thought not to exhibit a disk geometry due to unfavorable inclination (DBS97) but
beside exhibiting the dual-dust chemistry like CPD, its complex, multi-lobal geometry also
suggests binarity.

Infrared observations of HEN were obtained with VLT/NACO, VLTI/MIDI, VLT/ISAAC and TIMMI at
the ESO 3.6m. 
We attempted to detect and study small scales 
structures in the Mid-IR with the long baseline interferometer MIDI/VLTI. The object was over-resolved with 46m baselines and no interferometric data could be recorded, but the
acquisition images were very instructive.
Hen 2-113 exhibits a clear 1" torus-like structure
superimposed to a more diffuse environment visible in the L' (3.8$\mu$m), M' (4.78$\mu$m) and
8.7$\mu$m bands. Our ISAAC and TIMMI observations indicates that the PAHs are mostly concentrated towards the bipolar lobes of the nebula.

Photometry of the central star in L' and M' band indicates that it is $\sim$300 and $\sim$800
times brighter than predicted by stellar models (from De Marco et al. 1997) in $\rm L^\prime$ and $\rm M^\prime$ band
respectively. Moreover, the central object appears resolved in L' band with measured
FWHM of about 155\,mas. Simple calculations indicate that this infrared excess can be
explained by emission from hot dust grains. A mass of $\sim10^{-9}\rm
M_{\odot}$ with T$\sim$900K can account for this infrared excess. By a totally
independent way (fitting of the SED), Sahai et al. (2000) indicated the
possible presence of hot dust (T$\sim$900K and  $\rm M\sim 10^{-9}\rm
M_{\odot}$) inside the nebula, dust that we have apparently detected.
This is a clear evidence that dust is still continuously created in the present wind of the [WC] star.

\section{CPD-56$^\circ$8032}
De Marco, Barlow \& Cohen (2002), presented the first direct evidence for an edge-on disk/torus around CPD -568032, as revealed by recent HST/STIS spectroscopy.
Their HST spectra show a
spatially resolved continuum split into two bright peaks separated by 0".10
and interpreted to be stellar light reflected above and below an
obscuring dust disk. From these observations, the disk thickness of CPD is deduced to
be 134\,au, at 1.35 kpc (De Marco et al. 1997).

We observed CPD with VLTI/MIDI in imaging providing a typical 300\,mas resolution and in interferometric mode using UT2-UT3 47m baseline providing a typical spatial resolution of 20\,mas. We made also use of unpublished HST/WPC2 images in F435 and F606 filters.
The visible HST images exhibit a complex multilobal geometry dominated by a few faint lobes for which a counterpart is not always found. The farthest structures are located more than 6$\arcsec$ from the star. The mid-IR environment of CPD-568032 is dominated by a compact source, 
barely resolved by a single UT telescope in a 8.7$\mu$m filter ($\Delta \lambda$=1.6$\mu$m, contaminated by PAHs
emission).

The infrared core is almost fully
resolved with the three 40-45m projected baselines, ranging from -5$^\circ$ to 51$^\circ$, but smooth oscillating fringes at low level have been detected in dispersed visibilities. This clear signal is
interpreted in terms of a ring structure which would define the
bright inner rim of the equatorial disk. Geometric models allowed us to retrieve the main geometrical parameters of the disk. 

The object is proposed to be a central star (not visible in N band), that illuminates and heats a surrounding dusty disk, heavily resolved by the interferometer. The disk geometry consists in a circular bright inner rim and a rapidly decreasing, although important, flux in the outer regions, which appears to provide good  results. The first minimizations were done using a truncated circular Gaussian profile and a fixed inner gap radius equal to the FWHM of the Gaussian (real disk coupled). The residuals are surprisingly good but there are still important deviations from the measurements. The model has been refined with the incorporation of free inner gap radius and ellipticity, which matches the observed oscillations quite well.
For instance, a reasonably good fit is reached with an achromatic and elliptical truncated Gaussian with a radius of 97$\pm$11AU, an inclination of 28$\pm$7$^\circ$ and a PA angle for the major axis at -15$^\circ \pm$7$^\circ$.

\begin{figure}
  \begin{center}

\label{figvisi}
 \includegraphics[width=8.cm]{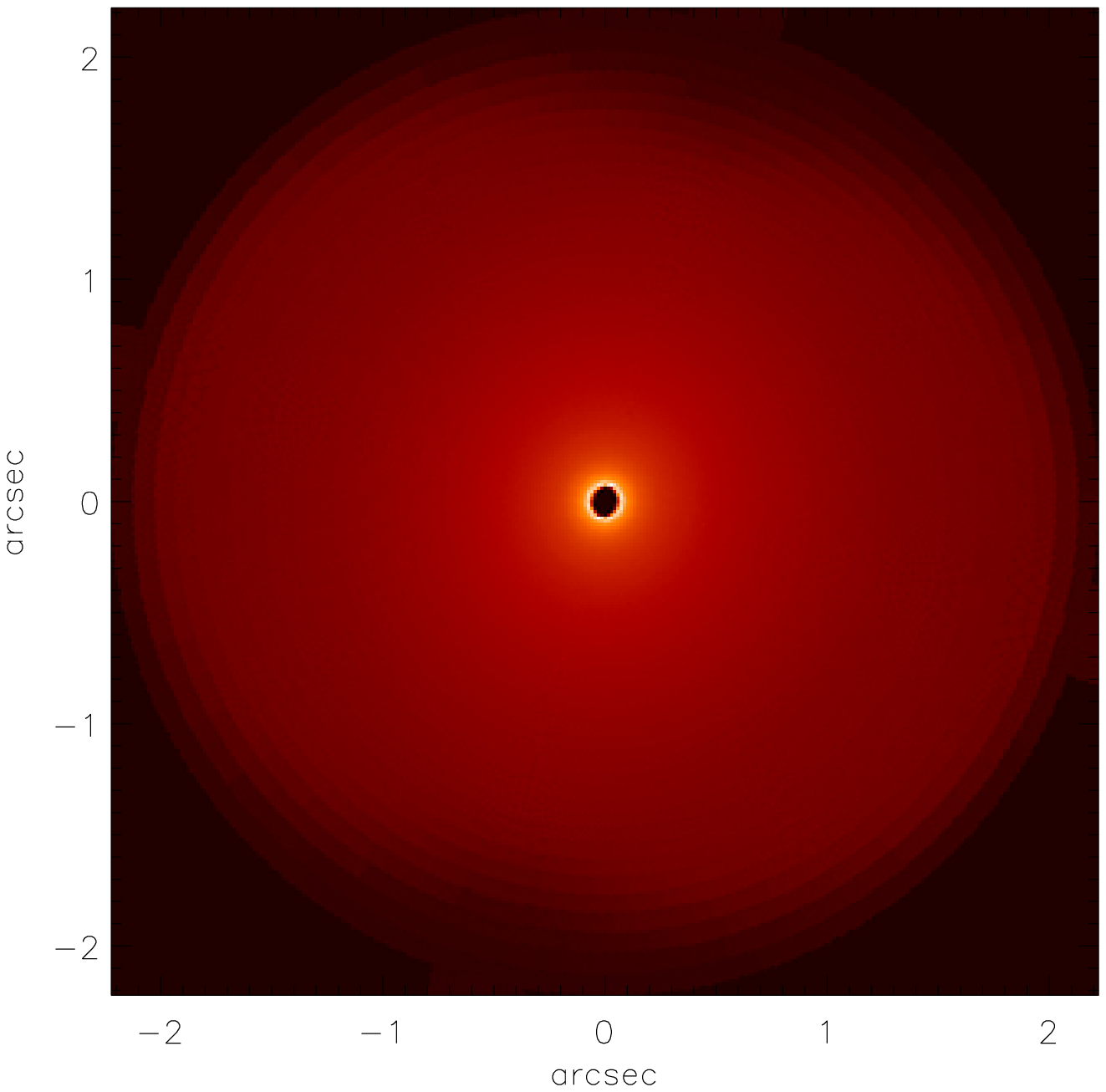}
 \includegraphics[width=4.cm, height=10.cm, angle=-90]{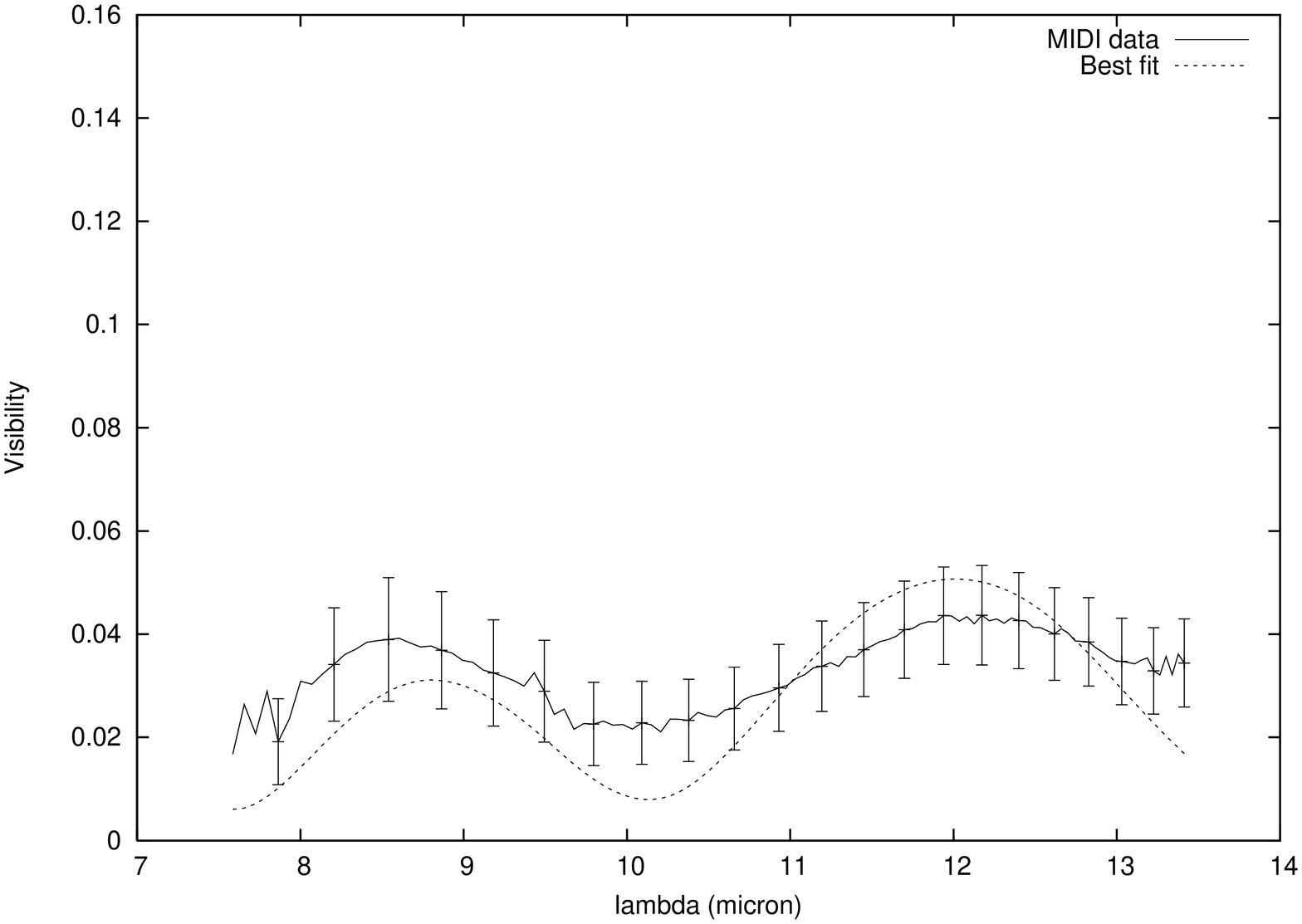}
\includegraphics[width=4.cm, height=10.cm,angle=-90]{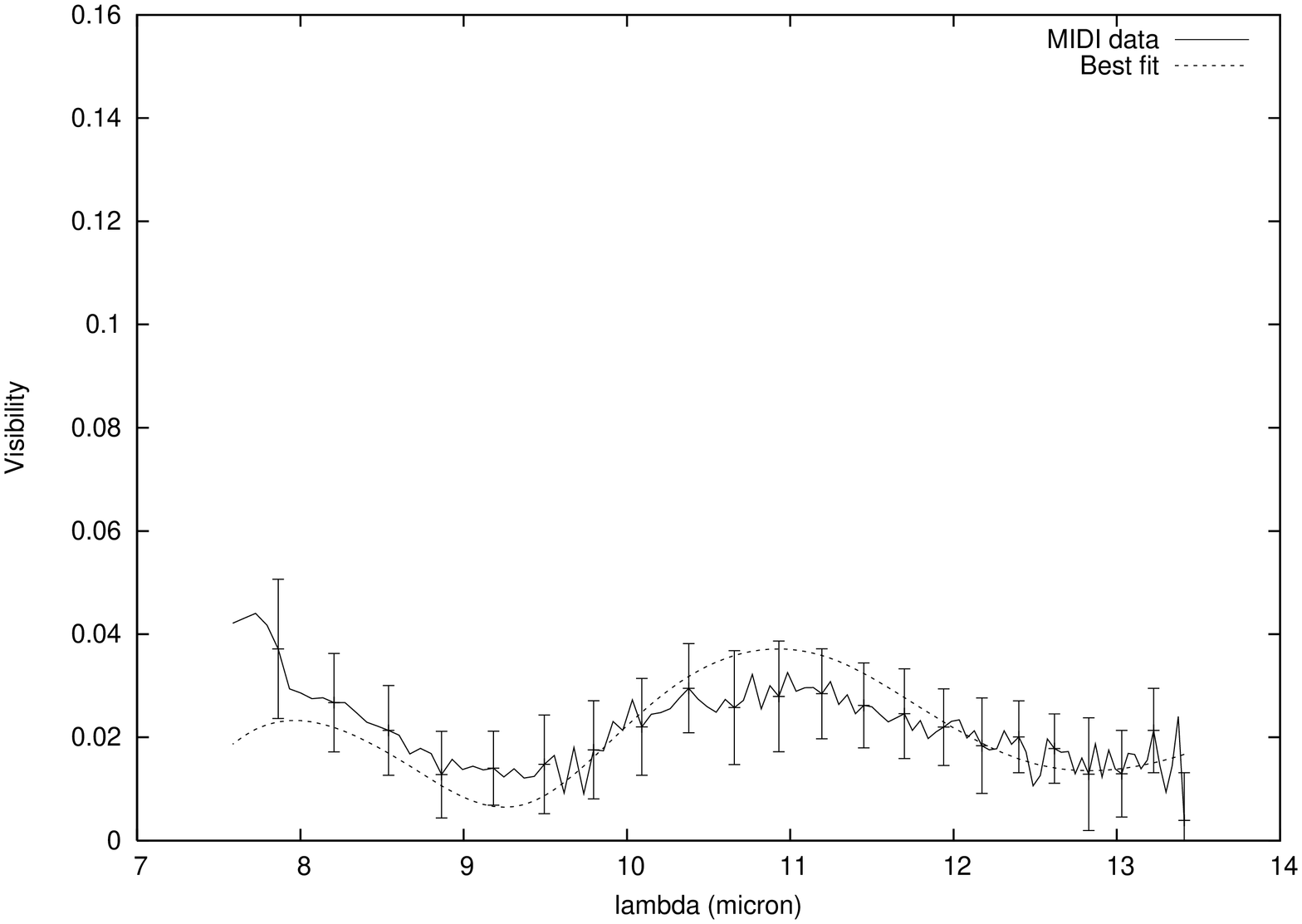}
\includegraphics[width=4.cm, height=10.cm, angle=-90]{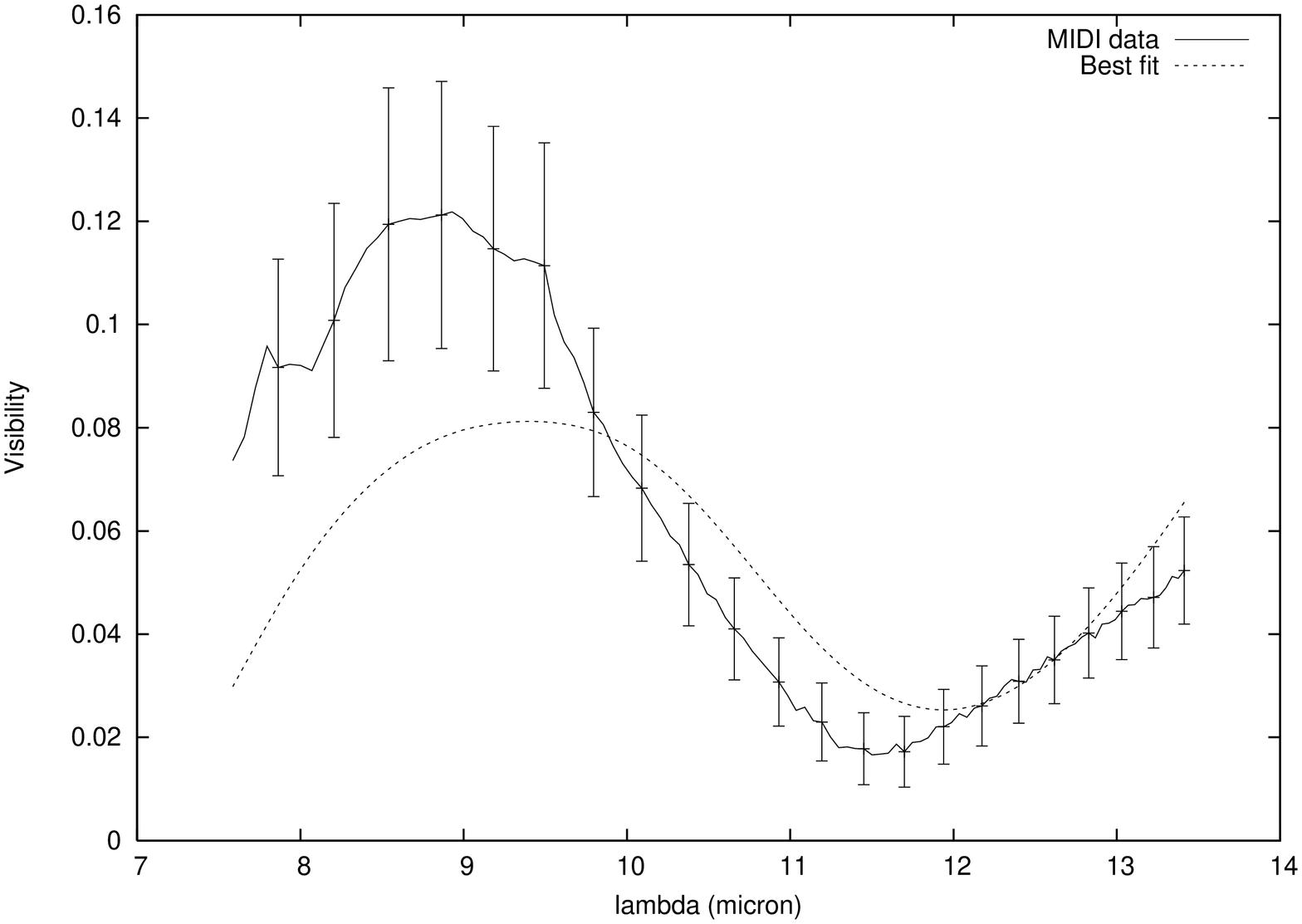}

  \caption{Up: the 10$\mu$m flux distribution of the radiative transfer model. Bottom: the three MIDI visibility measurements (solid lines) and the model signal (dotted lines). Most of the flux is over-resolved by the interferometer but the low level sinusoidal variation is the signature of the brigth inner rim of the disk. The two first visibility spectra have been obtained with almost identical projected baselines and a small variation of PA angle. The third one has been obtained with a smaller projected baseline with a clear change of PA angle allowing a good constraint of the aspect ratio of the disk.}
   \end{center}
\end{figure}

We performed radiative transfer simulations using the
continuum radiative transfer code MC3D (Wolf et al. 1999), restricted to carbon-rich dust (i.e. to the close vicinity of the star). The goal is to get a good estimate of the
physical parameters of the inner rim based on the simultaneous constraints of the
visibilities, MIDI spectra and the SED.

\section{A short comparison of these two environments}
HEN and CPD offer a unique opportunity to compare two different nebulae ejected from central stars presenting striking similarities. The central stars are both of [WC10] types, with very similar temperature, luminosity putting them into tracks of stars at 0.62M$_\odot$ (DBS97). At large scale, their environments observed by the HST are quite different. The nebula around HEN is more compact and perhaps simpler than the one seen around CPD although both exhibit a complex multi-lobal geometry. Despite of its compactness, HEN presents a ring-like structure which is resolved by a single 8m telescopes from L' to N band, whereas CPD is definitely only barely resolved at 8.7$\mu$m (probably due to the presence of PAHs) and harbor a compact disk in the vicinity of the central disk. Shortly speaking, it is tempting to propose that HEN exhibits a 'typical' torus formed in the post-AGB phase whereas the disk around CPD is more suggestive of a nebula evolution involving a binary companion.

In this context, the question of the dynamical age of the nebulae is of importance. The development of a compact and complex planetary nebula is a fast event compared to the life-time of their parent stars and it is possible that the differences reported for HEN and CPD nebulae are solely due to the age of the ejection, the nebula of CPD being younger in this case than HEN since the dusty core is more compact. In an other hand, it does not explain why some lobes seen in the HST images of CPD are so extended, reaching 6" from the central star, whereas the lobes from HEN are at most at 3" from the star.

Our high spatial resolution observations have considerably enlarged our knowledge of the dust geometry of these nebula but the key of the problem remain unanswered: is their geometry a consequence of binary interaction? In the case of CPD we could answer that the probability is large but in the case of HEN the evidence is less direct. There is a correlated issue: we have shown that the core of HEN's nebula is not a naked star, but that an optically thin shell of newly formed dust surrounds the star. The question is to know how this dust is formed at the vicinity of a [WR] star. Again, our view of this phenomenon would be clarified if we knew whether or not HEN harbors a binary system and of which kind. Understanding how dust can be formed in the hostile environment of hot winds is a challenging problem and a comparison with the massive WR stars case is instructive. 

Tuthill, Monnier, Danchi and collaborators have discovered that
two supposed isolated massive WR known as persistent dust producers, WR 104 and WR 98a stars were in fact
close-binary systems whose dust distribution was following the
orbital motion of the stars (Tuthill et al, 1999; Monnier et al,
1999, 2002). The dust production is probably triggered by the wind-wind
collision and these discoveries raised interesting questions and propositions
(see Monnier et al. 1999): is binarity mandatory to produce the dust?
Furthermore, are alla the WC8-9 spectral stars  a consequence of binary interaction?
These questions have strong resonances in the case of [WR] at the center of planetary nebulae for which the evolutionary status remains unclear and could also be the result of binary interaction. We note that in the frame of the scenario of De Marco \& Soker, the companion is no longer orbiting around the newly born [WR] star and that the dust production must in this context be formed from the wind alone.  

In the case of HEN, we did not detect any asymmetry which would suggest that the dust production occurs in a preferred plane but the spatial resolution provided by NACO in L' is probably too low. Moreover, the central object is no longer visible in MIDI N band observations and probably much below the sensitivity limit for interferometric detection.  The observations of the central core of HEN and CPD by AMBER in K band could be useful, but their magnitudes in this band (7.5 and 6.8 respectively) are demanding in regards to the instrument sensitivity. The observations of the core of such objects remains challenging.

\acknowledgements 
This work is the result of a fruitful collaborations involving Eric Lagadec, Arnaud Collioud, 
Orsola De Marco, Jos\'e de Freitas de Pacheco, Sebastian Wolf, Albert Zijlstra, 
Mikako Matsuura, Agn\`es Acker, Geof Clayton, Bruno Lopez and Alexander Rothkopf.


\end{document}